\documentclass[twocolumn,prl,floatfix,superscriptaddress]{revtex4-1}
\usepackage{color}
\usepackage{graphicx}
\usepackage{amsmath}
\usepackage{amssymb}
\usepackage{mathtools}
\usepackage{wasysym}
\usepackage{mathrsfs}
\usepackage[english]{babel}
\usepackage{url}

\newcommand{\be}{\begin{equation}}
\newcommand{\ee}{\end{equation}}
\newcommand{\nn}{\nonumber}
\newcommand{\trans}{T}
\graphicspath{{figures/}}

\begin{document}

\title{Fast machine-learning online optimization of ultra-cold-atom experiments}

\author{P.~B.~Wigley}
\affiliation{Quantum Sensors and Atomlaser Lab, Department of Quantum Science, Australian National University, Canberra, 2601, Australia}
\author{P.~J.~Everitt}
\affiliation{Quantum Sensors and Atomlaser Lab, Department of Quantum Science, Australian National University, Canberra, 2601, Australia}
\author{A.~van~den~Hengel}
\affiliation{Australian Centre for Robotic Vision, University of Adelaide, Adelaide, 5005, Australia}
\author{J.~W.~Bastian}
\affiliation{School of Computer Science, University of Adelaide, Adelaide, 5005, Australia}
\author{M.~A.~Sooriyabandara}
\affiliation{Quantum Sensors and Atomlaser Lab, Department of Quantum Science, Australian National University, Canberra, 2601, Australia}
\author{G.~D.~McDonald}
\affiliation{Quantum Sensors and Atomlaser Lab, Department of Quantum Science, Australian National University, Canberra, 2601, Australia}
\author{K.~S.~Hardman}
\affiliation{Quantum Sensors and Atomlaser Lab, Department of Quantum Science, Australian National University, Canberra, 2601, Australia}
\author{C.~D.~Quinlivan}
\affiliation{Quantum Sensors and Atomlaser Lab, Department of Quantum Science, Australian National University, Canberra, 2601, Australia}
\author{P.~Manju}
\affiliation{Quantum Sensors and Atomlaser Lab, Department of Quantum Science, Australian National University, Canberra, 2601, Australia}
\author{C.~C.~N.~Kuhn}
\affiliation{Quantum Sensors and Atomlaser Lab, Department of Quantum Science, Australian National University, Canberra, 2601, Australia}
\author{I.~R.~Petersen}
\affiliation{School of Engineering and Information Technology,
University of New South Wales at the Australian Defence Force Academy, Canberra, 2600, Australia}
\author{A.~Luiten}
\affiliation{Institute for Photonics \& Advanced Sensing, The School of Chemistry and Physics,The University of Adelaide, Adelaide, 5005, Australia}
\author{J.~J.~Hope}
\affiliation{Department of Quantum Science, Australian National University, Canberra, 2601, Australia}
\author{N.~P.~Robins}
\affiliation{Quantum Sensors and Atomlaser Lab, Department of Quantum Science, Australian National University, Canberra, 2601, Australia}
\author{M.~R.~Hush}
\email{M.Hush@adfa.edu.au}
\affiliation{School of Engineering and Information Technology,
University of New South Wales at the Australian Defence Force Academy, Canberra, 2600, Australia}

\date{\today}

\begin{abstract} 
Machine-designed control of complex devices or experiments can discover  strategies superior to those developed via simplified models.  We describe an online optimization algorithm based on Gaussian processes and apply it to optimization of the production of Bose-Einstein condensates (BEC).  BEC is typically created with an exponential evaporation ramp that is approximately optimal for s-wave, ergodic dynamics with two-body interactions and no other loss rates, but likely sub-optimal for many real experiments. Machine learning using a Gaussian process, in contrast, develops a statistical model of the relationship between the parameters it controls and the quality of the BEC produced.  This is an online process, and an active one, as the Gaussian process model updates on the basis of each subsequent experiment and proposes a new set of parameters as a result. We demonstrate that the Gaussian process machine learner is able to discover a ramp that produces high quality BECs in 10 times fewer iterations than a previously used online optimization technique. Furthermore, we show the internal model developed can be used to determine which parameters are essential in BEC creation and which are unimportant, providing insight into the optimization process.  
\end{abstract}

\maketitle

Experimental research into quantum phenomena often requires the optimization of resources or processes in the face of complex underlying dynamics and shifting environments. For example, creating large Bose-Einstein condensates (BECs) with short duty cycles is one of the keys to improving the sensitivity of cold-atom based sensors \cite{robins_atom_2013} or for performing scientific investigation into condensed matter phases \cite{bloch_quantum_2012}, many-body physics \cite{bloch_many-body_2008} and non-equilibrium dynamics \cite{langen_ultracold_2015}. The standard process of BEC production is evaporative cooling \cite{olson_optimizing_2013}; microscopic semi-classical theory exists to describe this process \cite{sackett_optimization_1997}, but it can oversimplify the dynamics and miss more complex and effective methods of performing evaporation. For example, Shobu \emph{et al.} \cite{shobu_optimized_2011} found circumventing higher order inelastic collisions can produce very large condensates. `Tricks' like this are likely to exist for other species with complicated scattering processes \cite{altin_collapse_2011}, but discovery is only possible by experimentation. We automate this process of discovery with \emph{machine-learning} online optimization (MLOO). What distinguishes our approach from previous methods for automation is that we seek to develop a statistical model of the relationship between parameters and the outcome of the experiment. We demonstrate that MLOO can discover condensation with less experiments than a competing optimization method and provide insight into which parameters are important in achieving condensation.

\begin{figure*}[t!hb]
\includegraphics[width=\textwidth]{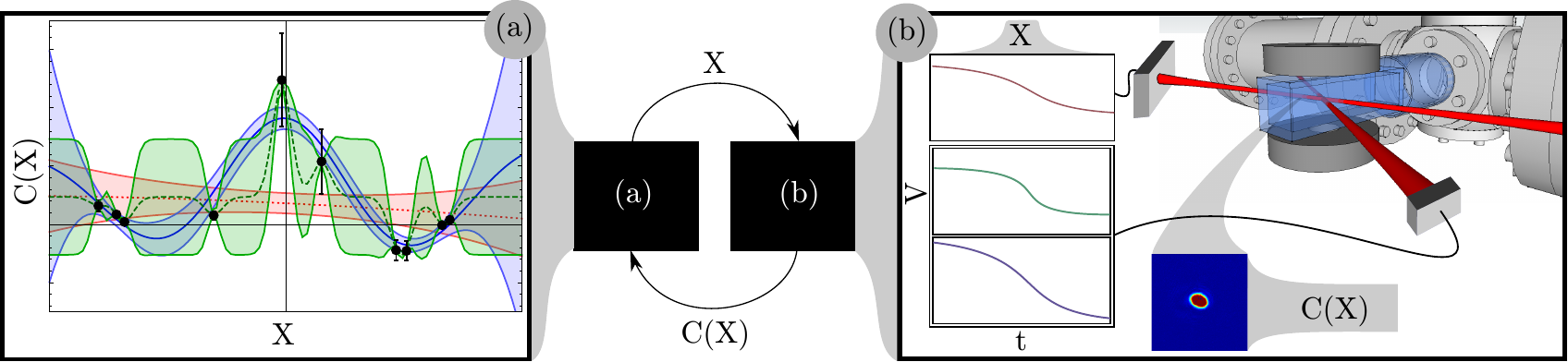}
\caption{The experiment and the `learner' see each other as a black box. The learner produces a parameter set, $X$, for the experiment to test, these are converted into cooling ramps and used to perform an experiment. After the evaporation process is finished, an image of the cold atoms taken is used to calculate a cost function based on its quality as a resource $C(X)$. $C(X)$ is then fed back to the learner. The learner uses a GP to model the relationship between the input parameter values and the cost function values produced by the experiment. This model depends on a set of correlation lengths, or hyperparameters. Part (a) of the figure plots a set of observed costs (black circles with bars for uncertainty) with three possible GP models fit to the data: one with a long correlation length (red dotted), a medium correlation (blue solid) and a short correlation length (green dashed).  Each GP is illustrated by a mean cost function bracketed by two curves indicating the function +/- one standard deviation from the mean.  The correlations lengths affect both the mean and uncertainty of the model; note that the uncertainty approaches zero near the observed points. A final cost function is then produced as a weighted average over the most likely correlation lengths. This predicted model is then used to pick the next parameter set to test $X$, which is fed to the machine.}
\end{figure*}

Online optimization (OO), with mostly genetic \cite{judson_teaching_1992,warren_coherent_1993,amstrup_genetic_1995,dods_genetic_1996,baumert_femtosecond_1997,pearson_coherent_2001,zeidler_evolutionary_2001,walmsley_quantum_2003,rohringer_stochastic_2008,tsubouchi_rovibrational_2008,rohringer_stochastic_2011,starkey_scripted_2013} but also gradient \cite{roslund_gradient_2009} and hybrid solvers \cite{egger_adaptive_2014,kelly_optimal_2014}, has been used to enhance a variety of quantum experiments. Here, online means optimization is done in real time with the experiment. We apply OO based on machine learning. What distinguishes our approach is that it does not only seek to optimize the experiment, but also creates an internal model that is able to predict the performance of future experiments given any set of parameters.  This is achieved by modeling the experiment using a Gaussian process (GP) \cite{rasmussen_gaussian_2006}. Online machine learning (OML) with GPs \cite{seo_gaussian_2000,csato_sparse_2002,rasmussen_gaussian_2006,deisenroth_gaussian_2009,gramacy_particle_2011} has been applied in a variety of areas including robotics \cite{nguyen-tuong_local_2008,nguyen-tuong_local_2009}, vision \cite{ranganathan_online_2011}, industrial chemistry \cite{yu_online_2012,li_multiple_2009} and biochemistry \cite{gao_gaussian_2008}. However, in all of these cases, the focus was not on optimization. Rather, the goal was the development of an accurate model. We here combine the advantages of OML with the motivation of OO. The resultant MLOO algorithm has the following beneficial features: every experimental observation from an optimization run is used to improve the GP model, and uncertainties in the measurements can be correctly accounted for; our algorithm is both deterministic and able to find global minima, instead of randomly exploring the parameter space our learner uses its estimate of the variance from the GP to pick parameter values where we are most uncertain about the value of the cost function; fast gradient-based optimization routines can not be applied directly to our experimental data as it is too noisy, but they can be employed to efficiently find optima in our smooth machine-generated model;  estimation of parameter sensitivity and visualization of the functional dependence of the resource's quality can inform experimentalists on how to best develop future optimization experiments. 

The experimental apparatus is described in detail in \cite{kuhn_Bose-condensed_2014}. Initially $^{87}$Rb atoms are cooled in a combined 2D and 3D MOT system and subsequently cooled further by RF (radio frequency) evaporation. The cloud is then loaded into a cross beam optical dipole trap for the final evaporation stage. It is this stage that is the subject of the optimization process. The cross dipole trap is formed from two intersecting $1090$nm and $1064$nm lasers with approximate waists of $100\mu$m and $60\mu$m respectively. The depth of the cross trap is determined by the intensity of the two beams. The $1064$nm beam is controlled by varying the current to the laser, while the $1090$nm beam is controlled using the current and a waveplate rotation stage combined with a polarizing beamsplitter to provide additional power attenuation while maintaining mode stability. A diagram of the experimental set up is shown in figure 1. Normally the power to these beams is ramped down over time, thereby lowering the walls of the trap and allowing the higher energy atoms to leak out. The remaining atoms rethermalize to a lower temperature, enabling cooling. Once the gas has been cooled to temperatures on the order of nK, a phase transition occurs, and a macroscopic number of atoms start to occupy the same quantum state. This transition is called Bose-Einstein condensation \cite{anglin_boseeinstein_2002}. We hand over control of these ramps to the MLOO. We consider two parameterizations: one simple, where we only control the start and end points of a linear interpolation; and one complex, where we add variable quadratic, cubic and quartic corrections to the simple case (see Appendix).

The approach we propose is a form of supervised learning, meaning that we provide the learner with a number that quantifies the quality of the resource produced or in optimization terminology a cost that must be minimized. Na\"{i}vely one might try to use a measure based on temperature and particle number. However determining these quantities accurately near condensation is difficult when constrained to very few runs per parameter set. Instead, a semi-heuristic method is used to calculate the cost. An absorption image of the final state of the quantum gas is taken after a 30ms expansion of the cloud, with the image providing the optical depth as a function of space. The cost is calculated from all data between a lower and upper threshold optical depth. The lower threshold is determined by the noise in the system. The upper threshold is required since the absorption images are taken on resonance and hence saturate for areas with very large optical depth. The upper threshold is set slightly below this saturation level. Only data from between the bounds is used and cost is simply the average of these values. In practice this means the sharper the edges of the cloud, the lower the cost. Indeed, low quality thermal clouds have broad edges, whereas the ideal BEC has much sharper edges. Each parameter set is tested twice with the average of the two runs used for the cost. Tests of the variation in cost for a set of parameters run-to-run indicate they obey a Gaussian distribution. As such we are able to estimate the uncertainty from two runs as twice the range. In doing so, the chance we have underestimated the uncertainty will be 27\%. We therefore also apply bounds to the uncertainty to eliminate outliers overly affecting the modeling process. The cost function can be evaluated as long as some atoms are present at the end of the evaporation run. In cases where the evaporation parameters produced no cloud twice for a set of parameters, we set the cost to a default high value.

We treat the experiment as a stochastic process $\mathscr{C}(X)$ which is dependent on the parameters $X = (x_1,\cdots,x_M)$. When we make a measurement and determine a cost, we interpret this as a sample of this process $C(X)$ with some associated uncertainty $U(X)$. We define the set of all parameters, costs and uncertainty previously measured as $\mathcal{X} = (X_1,\cdots X_N)$, $\mathcal{C} = (C_1,\cdots, C_N)$ and $\mathcal{U} = (U_1,\cdots, U_N)$ respectively and collectively refer to these sets as our observations $\mathcal{O} = (\mathcal{X},\mathcal{C},\mathcal{U})$. The aim of OO is to use previous observations $\mathcal{O}$ to plan future experiments in order to find a set of parameters that minimize the mean cost of the stochastic process $M_{\mathscr{C}}(X)$. Unique to the MLOO approach, we first make an \emph{estimate} of the stochastic process given our observations $\hat{\mathscr{C}}(X|\mathcal{O})$, which is then used to determine what parameters to try next.

We model $\mathscr{C}(X)$ as a GP --- a distribution over \emph{functions} --- with constant mean function and covariance defined by a squared exponential correlation function $K(X,X',H) = e^{-\sum_{j=1}^M (x_j- x_j')^2/h_j^2}$ where $H = (h_1, \cdots, h_M)$ is a set of correlations lengths for each of the parameters. The mean function conditional on the observations $\mathcal{O}$  and correlations lengths $H$ of our GP is: $\mu_{\hat{\mathscr{C}}}(X|\mathcal{O}, H)$, which is evaluated through a set of matrix operations \cite{rasmussen_gaussian_2006} (see Appendix). As we are using a GP, we can also get the variance of the functions conditioned on $\mathcal{O}$ and $H$: $\sigma_{\hat{\mathscr{C}}}(X|\mathcal{O}, H)$ \cite{rasmussen_gaussian_2006}. Both of these estimates depend on the correlation lengths $H$, normally referred to as the \emph{hyperparameters} of our estimate. We assume that $H$ is not known a priori and needs to be fitted online.

The correlation lengths $H$ control the sensitivity of the model to each of the parameters, and relates to how much a parameter needs to be changed before it has a significant effect on the cost (see Fig.~1). A standard approach to fit $H$ is maximum likelihood estimation \cite{rasmussen_gaussian_2006}. Here, the hyperparameters are globally optimized over the likelihood of the parameters $H$ given our observations $\mathcal{O}$, or $L(H | \mathcal{O})$ \cite{rasmussen_gaussian_2006} (see Appendix). However, when the data set is small there will often be multiple local optima for the hyperparameters whose likelihoods are comparable to the maximum. We term these hyperparameters the hypothesis set $\mathcal{H}= (H_1,\cdots,H_P)$ with corresponding likelihood set $\mathcal{L} = (L_1,\cdots,L_P)$. 

To produce our final estimates for the mean function and variance we treat each hypothesis as a \emph{particle} \cite{gramacy_particle_2011}, and perform a weighted average over $\mathcal{H}$. The weighted mean function is now defined as $M_{\hat{\mathscr{C}}}(X|\mathcal{O}, \mathcal{H}) \equiv \sum_{i=1}^P w_i \mu_{\hat{\mathscr{C}}}(X|\mathcal{O}, H_i)$ and weighted variance of the functions is $\Sigma_{\hat{\mathscr{C}}}^2(X|\mathcal{O}, \mathcal{H}) \equiv \sum_{i=1}^P w_i(\sigma_{\hat{\mathscr{C}}}^2(X|\mathcal{O}, H_i) + \mu_{\hat{\mathscr{C}}}^2(X|\mathcal{O}, H_i)) - M_{\hat{\mathscr{C}}}^2(X|\mathcal{O}, \mathcal{H})$, where $w_i = L_i/\sum_{i=1}^P L_i$ are the relative weights for the hyperparameters. Now that we have our final estimate for $\mathscr{C}(X|\mathcal{O}, \mathcal{H})$, we need to determine an optimization \emph{strategy} for picking the next set of parameters to test. 

\begin{figure}[t!]
\includegraphics[width=\columnwidth]{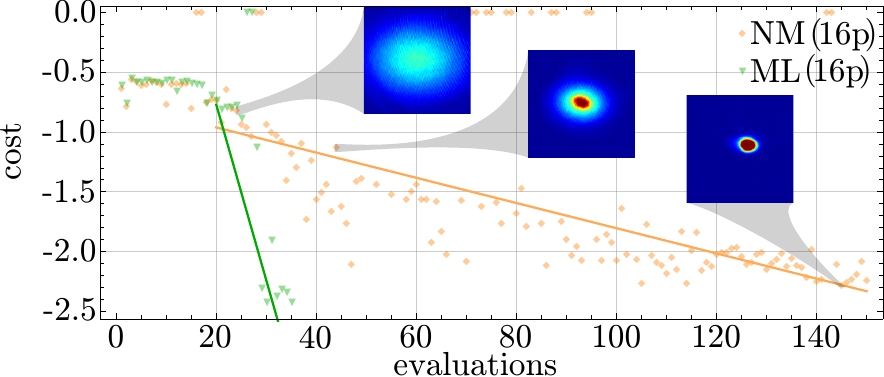}
\caption{The optimization of the evaporation stage of creating a BEC using the complex 16 parameter scheme. The first 20 evaluations are an initial training run using a simple Nelder-Mead algorithm. The machine learning algorithm (green) then quickly optimizes to BEC. The insets show the different regimes that the experiment goes through, from a large, completely thermal cloud through to the sharp edged BEC.}
\end{figure}

Consider the following homogeneous strategies: we could test parameters that minimize $M_{\hat{\mathscr{C}}}(X)$, but this strategy can get trapped in local minima; or we could test parameters that maximize $\Sigma_{\hat{\mathscr{C}}}^2(X)$ (i.e. where we are most uncertain), but this may require a large number of trials to map the space and would not prioritize refinement of the minima. We chose to implement an inhomogeneous strategy that repeatably \emph{sweeps} between these two extremes by minimizing a biased cost function: $B_{\hat{\mathscr{C}}}(X) \equiv b M_{\hat{\mathscr{C}}}(X) - (1 - b) \Sigma_{\hat{\mathscr{C}}}(X)$, where the value for $b$ is linearly increased from $0$ to $1$ in a cycle of length $Q$. This makes the learner change strategy from getting maximum information ($b=0$), to looking for a new minima: with a high risk-seeking ($b$ small) to risk-neutral ($b=1$) preference. During testing with synthetic data, we found sweeping was more robust and efficient than the homogeneous approaches. When we minimized $B_{\hat{\mathscr{C}}}(X)$ we also put bounds, set to 20\% of the parameters maximum-minimum values, on the search relative to the last best measured $X$. We call these bounds a \emph{leash}, as it restricts how fast the learner could change the parameters but did not stop it from exploring the full space (similar to trust-regions \cite{conn_trust_2000, yuan_review_2000}). This was a technical requirement for our experiment: when a set of parameters was tested that was very different from the last set, the experiment almost always produced no atoms, meaning we had to assign a default cost that did not provide meaningful gradient information to the learner. Once the next set of parameters is determined they are sent to the experiment to be tested. After the resultant cost is measured this is then added to the observation set $\mathcal{O}$ with $N\rightarrow N+1$ and the entire process is repeated. 

We emphasize that fitting of $H$, estimation of $M_{\hat{\mathscr{C}}}(X)$ and minimization of $B_{\hat{\mathscr{C}}}(X)$ is all done online while the experiment is being run. In a single optimization run, the learner typically performs hundreds more hypothetical experiments than the number physically run in the lab. The MLOO algorithm we developed is open source and available online at \cite{hush_m-loop_2015} (it uses the package scikit-learn \cite{pedregosa_scikit-learn:_2011} to evaluate the GPs). 

As a benchmark for comparison, we also performed OO using a Nelder-Mead solver \cite{nelder_simplex_1965}, which has previously been used to optimize quantum gates \cite{kelly_optimal_2014}.  

We demonstrate the performance of machine learning online optimization in comparison to the Nelder-Mead optimizer in Fig.~2. Here we used the complex parameterization for all 3 ramps, and added an extra parameter that controlled the total time of the ramps, resulting in 16 parameters. If we were to perform a brute force search and optimize the parameters to within a $10\%$ accuracy of the parameters maximum-minimum bounds, the number of runs required would be $10^{16}$. The Nelder-Mead algorithm is able to find BEC much faster than this, in only 145 runs. The machine learning algorithm, on the other hand, is much faster. After the first 20 training runs, where the machine learning and Nelder-Mead algorithm use a common set of parameters, the machine learning algorithm converges in only 10 experiments. 

\begin{figure}[t!]
\includegraphics[width=\columnwidth]{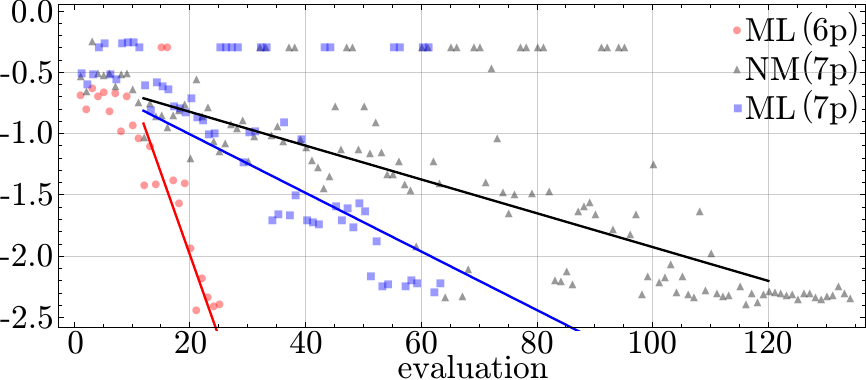}
\caption{Optimization of evaporation curves to produce a BEC. The first 2N evaluations use a simple Nelder-Mead algorithm to learn about the cost space. The machine learning algorithm (red and blue) optimizes to BEC faster than the Nelder-Mead (black). By utilizing the machine learning model a parameter is eliminated and the convergence improves (red).}
\end{figure}

The learner used in Fig.~2 only used the \emph{best} hypothesis set when picking the next parameters, in other words we set $P=1$. Evaluating multiple GPs is computationally expensive with so many parameters, so to save time we made this restriction. In spite of this, the learner discovered ramps that produced BEC in very few iterations. This is because the learner consistently fitted the correlation lengths of the 3 most important parameters --- the end points of the ramps --- very quickly. However, we found the other correlations lengths were not estimated well and would not converge, even after a BEC was found. This meant that we were unable to make useful predictions about the cost landscape and we could not reliably determine what parameters were least important. 

Gramacy \emph{et al.} \cite{gramacy_particle_2011} have suggested that making good online estimation of the GP correlation lengths requires multiple particles. We considered achieving this goal in a different experiment as shown in Fig.~3. Here we used a learner with many particles $P=16$, but had to use the simple parameterization for the ramps to save computational time. This resulted in a total of 7 parameters. We can see again the overall trend for the machine learner is still faster than Nelder-Mead, but less pronounced. More carefully estimating the correlation lengths has hindered the convergence rate compared to the $16$ parameter case. Nevertheless, as we now have a more reliable estimate of the correlation lengths we can take advantage of a different feature of the learner.

In Fig.~4(a) we show estimates of the cost landscape as 1D cross sections about the best measured point. We plot the two most sensitive parameters and the least. We can see the least sensitive parameter appears to have no effect on the production of a BEC. This parameter corresponds to an intentionally added 7th parameter of the system that controls nothing in the experiment. Fig.~4(a) shows the learner successfully identified this, even with such a small data set. After making this observation we can then reconsider the design of the optimization process and eliminate this parameter from the experiment. 

\begin{figure}[t!]
\includegraphics[width=\columnwidth]{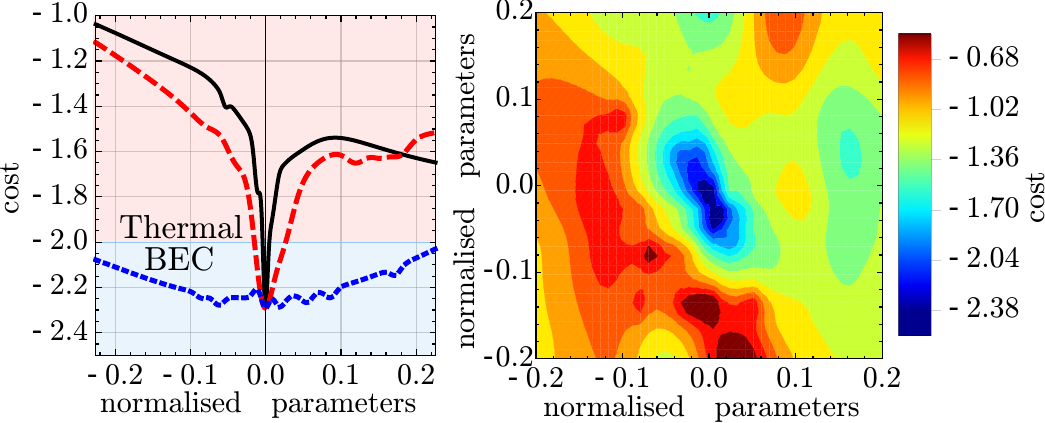}
\caption{Plots of the cross sections through the minima of the cost landscape as predicted by the learner. In (a) the predicted cost is shown as a function of the end of the polarization ramp (red), the end of the dipole beam ramp (green) and the unconnected parameter (blue). The learner correctly identifies that the unconnected parameter does not have a significant effect on the production of BEC. In (b) a cross section of the 2 most sensitive parameters are plotted against cost.}
\end{figure}

In Fig.~3 we plot the machine learner optimization run with $P=16$ but now with only 6 parameters. We can see the learner converges much more rapidly than the 7 parameter case, and even produces a higher quality BEC. As the learner no longer takes extra runs to determine the importance of the useless 7th parameter, it achieves BEC quite rapidly. Lastly, to give us some understanding of the complexity of the landscape, we plot a 2D cross section of the landscape against the two most sensitive parameters in Fig.~4(b) generated from the 6 parameter machine learning run. We can see there is a very sharp transition to BEC, as it exists in a very deep valley of the landscape. 

We have demonstrated that MLOO can discover evaporation ramps that produce high quality BECs with far fewer experiments than OO with Nelder-Mead. Rapid optimization of ultra-cold-atom experiments is not just a useful tool to overcome technical difficulties, but will be vital in the application of BECs in proposed space-based scientific investigations \cite{zoest_bose-einstein_2010,arimondo_atom_2009}. Furthermore, when implemented with many particles, the learner can be used to make estimates of the cost landscape to determine what parameters most contributed to BEC production and aid better experimental design. In future work we will apply MLOO to atomic species with more exotic scattering properties \cite{altin_collapse_2011} in order to find novel cooling ramps that find optimal solutions in the competition between poorly characterized complex dynamical processes. Our approach is generic and available \cite{hush_m-loop_2015} for use on other scientific experiments, ultra-cold atom based or otherwise. 


\begin{acknowledgments}
MRH acknowledges funding from an Australian Research Council (ARC) Discovery Project (project number DP140101779). JJH acknowledges support of an ARC Future Fellowship (FT120100291). AL would like to thank the South Australian Government through the Premier's Science and Research Fund for supporting this work.
\end{acknowledgments}

\bibliographystyle{apsrev4-1}
\bibliography{machinelearning}

\section{Appendix}

\noindent
\textit{Gaussian process evaluation:} In practice, evaluating a Gaussian process (GP) reduces to a set of matrix operations whose derivation is given by Rasmussen \emph{et al.} \cite{rasmussen_gaussian_2006} in section 2.7. Consider $N$ previous experiments have been performed with parameter sets $\mathcal{X} = (X_1,\cdots X_N)$ (each $X_j = (x_{1,j}, \cdots x_{M,j})$), measured costs $\mathcal{C} = (C_1,\cdots, C_N)$ and uncertainties $\mathcal{U} = (U_1,\cdots, U_N)$. We refer to the set of this data as our observations $\mathcal{O} = (\mathcal{X},\mathcal{C},\mathcal{U})$. We fit a GP to these observations with constant function offset $\beta$ and covariance defined by a squared exponential correlation function $K(X_p,X_q,H) = e^{-\sum_{j=1}^M (x_{j,p}- x_{j,q})^2/h_j^2}$ where $H = (h_1, \cdots, h_M)$ are the hyperparameters of the model. 

The mean function and variance of the functions are:
\begin{align}
\mu_{\hat{\mathscr{C}}}(X|\mathcal{O},H) = & \beta + r(X)^\trans \gamma  \\
\sigma_{\hat{\mathscr{C}}}^2 (X|\mathcal{O},H) = & \sigma_{\mathcal{C}}^2 ( 1 - r(X)^T R^{-1} r(X) \nn \\
& + (j^T R^{-1} j)^{-1} (j^\trans R^{-1} r(X) - 1)^2 ) \label{eqn:varC}
\end{align}
where $\sigma_{\mathcal{C}}^2$ is the variance of the costs $\mathcal{C}$, and we define the constant $\beta \equiv (j^\trans R^{-1} j)^{-1} j^\trans R^{-1} Y$, the  
$N \times 1$ vector $r(X)$ such that $\{r(X)\}_{1,i} = K(X,X_i,H)$, the $N \times 1$ vector $\gamma \equiv R^{-1}(Y - j \beta)$, the $N \times 1$ vector $Y$ of the costs defined by $\{Y\}_{1,i} = C_i$, the $N \times 1$ vector $\{j\}_{1,i} = 1$, the $N\times N$ matrix $R$ defined as $\{R\}_{i,j} = K(X_i,X_j,H) + \delta_{i,j} U_i^2$, and where
$\delta_{i,j}$ is the Kronecker delta function.  $\{\cdot\}_{i,j}$ is our notation for the $i$th row and $j$th column of a matrix or vector.

When finding the most likely hyperparameters we maximize the likelihood function. The likelihood $L(H | \mathcal{O})$ is defined as the probability of the costs given the parameters, uncertainties and hyperparameters: $P(\mathcal{C}| \mathcal{X},\mathcal{U},H)$, the log of which is:
\begin{align}
\log P(\mathcal{C}| \mathcal{X},\mathcal{U},H) = & \frac{1}{2}( - \log |R| - \log  j^\trans R^{-1} j \nn \\ 
& - (N - 1) \log 2\pi \nn - Y^\trans (R^{-1} \\
& - (j^\trans R^{-1} j)^{-1} R^{-1} j j^\trans R^{-1}) Y)  
\end{align}

\noindent
\textit{Parameterizations of evaporation ramps:} The simple parameterization of the evaporation ramps is
\begin{align}
\mathcal{R}_{s}(y_{i},y_{f},t_{f}) = y_{i}+\left(y_{f}-y_{i}\right)\frac{t}{t_{f}}
\end{align}
where $y_i$ and $y_f$ specify the start and end amplitudes of the ramps and $t_f$ specifies the length in time. 

The complex parameterization an extension of the simple form: 
\begin{align}
\mathcal{R}_{c}&(y_{i},y_{f},A_{1},A_{2},A_{3},t_{f}) = \nn \\
& y_{i}+\left(y_{f}-y_{i}\right)\frac{t}{t_{f}}+A_{2}t\left(t-t_{f}\right) \nn \\
   & +A_{3}t\left(t-t_{f}\right)\left(t+\frac{1}{2}t_{f}\right) \nn \\
   & +A_{4}t\left(t-t_{f}\right)\left(t+\frac{2}{3}t_{f}\right)\left(t+\frac{1}{3}t_{f}\right)
\end{align}
where $A_1$, $A_2$ and $A_3$ correspond to the 3rd, 4th and 5th order polynomial terms respectively with each polynomial having evenly spaced roots between $t=0$ and $t=t_f$. As with the simple parametrization $t_f$ specifies the end of the ramps in time.

In each of the three ramps being optimized, the parameters $y_{i}$, $y_{f}$, $A_{1}$, $A_{2}$, $A_{3}$ are independent. However, the final time $t_f$ is common. 

\end{document}